\documentclass[acmsmall]{acmart}

\AtBeginDocument{%
  \providecommand\BibTeX{{%
    \normalfont B\kern-0.5em{\scshape i\kern-0.25em b}\kern-0.8em\TeX}}}

\usepackage{listings}
\usepackage{xcolor}
\usepackage{diagbox}
\usepackage{makecell}
\definecolor{pblue}{rgb}{0.13,0.13,1}
\definecolor{pgreen}{rgb}{0,0.5,0}
\definecolor{pred}{rgb}{0.9,0,0}
\definecolor{pgrey}{rgb}{0.46,0.45,0.48}
\definecolor{applegreen}{rgb}{0, 0.5, 0.0}
\usepackage[flushleft]{threeparttable}
\usepackage{tabularx}
\usepackage{makecell}
\usepackage{tcolorbox}
\usepackage{multirow}
\usepackage{array}
\usepackage{enumitem}
\usepackage{graphicx}
\usepackage{soul}
\usepackage{tabularx}
\usepackage{makecell}
\usepackage{mdframed}
\definecolor{codeblue}{RGB}{20,76,134}
\definecolor{codegreysh}{RGB}{114,136,223}
\definecolor{code}{RGB}{51,51,255}
\lstdefinestyle{listingstyle}{
    language=Java,
    basicstyle=\ttfamily\scriptsize,
    keywordstyle=\bf\ttfamily\color{codeblue},
    stringstyle=\color{codegreysh},
    moredelim=[l][\bf\ttfamily\color{red}]{///},
    moredelim=[l][\bf\ttfamily\color{orange}]{//,},
    moredelim=[s][\bf\ttfamily\color{code}]{/**}{**/}
}

\setcopyright{acmlicensed}
\acmDOI{10.1145/3660773}
\acmYear{2024}
\copyrightyear{2024}
\acmSubmissionID{fse24main-p76-p}
\acmJournal{PACMSE}
\acmVolume{1}
\acmNumber{FSE}
\acmArticle{66}
\acmMonth{7}
\received{2023-09-28}
\received[accepted]{2024-04-16}

\begin{document}

\title{A Deep Dive into Large Language Models for Automated Bug Localization and Repair}

\author{Soneya Binta Hossain}
\orcid{0000-0002-7282-061X}
\affiliation{%
  \institution{University of Virginia}
  \city{Charlottesville}
  \country{USA}
}
\email{sh7hv@virginia.edu}

\author{Nan Jiang}
\orcid{0000-0001-8518-2576}
\affiliation{%
  \institution{Purdue University}
  \city{West Lafayette}
  \country{USA}
}
\email{jiang719@purdue.edu}

\author{Qiang Zhou}
\orcid{0009-0002-6770-0880}
\affiliation{%
  \institution{Amazon Web Services}
  \city{Santa Clara}
  \country{USA}
}
\email{zhouqia@amazon.com}

\author{Xiaopeng LI}
\orcid{0000-0003-4916-1131}
\affiliation{%
  \institution{Amazon Web Services}
  \city{Santa Clara}
  \country{USA}
}
\email{xiaopel@amazon.com}

\author{Wen-Hao Chiang}
\orcid{0000-0003-2300-738X}
\affiliation{%
  \institution{Amazon Web Services}
  \city{Santa Clara}
  \country{USA}
}
\email{cwenhao@amazon.com}

\author{Yingjun Lyu}
\orcid{0000-0002-0139-8028}
\affiliation{%
  \institution{Amazon Web Services}
  \city{Santa Clara}
  \country{USA}
}
\email{yingjunl@amazon.com}

\author{Hoan Nguyen}
\orcid{0000-0002-6194-7930}
\affiliation{%
  \institution{Amazon Web Services}
  \city{Santa Clara}
  \country{USA}
}
\email{hoanamzn@amazon.com}

\author{Omer Tripp}
\orcid{0000-0002-2393-854X}
\affiliation{%
  \institution{Amazon Web Services}
  \city{Santa Clara}
  \country{USA}
}
\email{omertrip@amazon.com}

\renewcommand{\shortauthors}{Hossain et al.}

\newcommand{\todoc}[2]{{\textcolor{#1}{#2}}}
\newcommand{\todoblack}[1]{{\todoc{black}{\textbf{[[#1]]}}}}
\newcommand{\todored}[1]{{\todoc{red}{\textbf{[[#1]]}}}}
\newcommand{\todogreen}[1]{\todoc{green}{\textbf{[[#1]]}}}
\newcommand{\todoblue}[1]{\todoc{black}{#1}}
\newcommand{\todoorange}[1]{\todoc{orange}{\textbf{[[#1]]}}}
\newcommand{\todobrown}[1]{\todoc{brown}{\textbf{[[#1]]}}}
\newcommand{\todogray}[1]{\todoc{gray}{\textbf{[[#1]]}}}
\newcommand{\todopurple}[1]{\todoc{purple}{\textbf{[[#1]]}}}
\newcommand{\todopink}[1]{\todoc{magenta}{\textbf{[[#1]]}}}
\newcommand{\todocyan}[1]{\todoc{cyan}{\textbf{[[#1]]}}}
\newcommand{\todoviolet}[1]{\todoc{violet}{\textbf{[[#1]]}}}
\newcommand{\todo}[1]{\todored{TODO: #1}}

\newcommand{\nan}[1]{\todopink{Nan: #1}}
\newcommand{\soneya}[1]{\todogreen{Soneya: #1}}
\newcommand{\qiang}[1]{\todoorange{Qiang: #1}}
\newcommand{\mr}[1]{\todoblue{#1}}
\newcommand{\toolname}{Toggle}



\newcommand{\code}[1]{\texttt{\small #1}} 

\newcounter{finding}
\newcommand{\finding}[1]{\refstepcounter{finding}
    \begin{mdframed}[linecolor=gray,roundcorner=12pt,backgroundcolor=gray!15,linewidth=3pt,innerleftmargin=2pt, leftmargin=0cm,rightmargin=0cm,topline=false,bottomline=false,rightline=false]
        \textbf{Finding \arabic{finding}:} #1
    \end{mdframed}
}

\setlist[itemize]{leftmargin=20pt}

\begin{abstract}
Large language models (LLMs) have shown impressive effectiveness  in various software  engineering tasks, including automated program repair (APR). In this study, we take a deep dive into automated bug localization and repair utilizing LLMs. \mr{In contrast to many deep learning-based APR methods that assume known bug locations, rely on line-level localization tools, or address bug prediction and fixing in one step, our approach uniquely employs LLMs to predict bug location at the token level and subsequently utilizes them for bug fixing. This methodological separation of bug localization and fixing using different LLMs enables effective integration of diverse contextual information and improved incorporation of inductive biases}. We introduce \textbf{\toolname: \underline{To}ken-\underline{G}ranulated Bug \underline{L}ocalization and R\underline{e}pair}, a comprehensive program repair framework that integrates a bug localization model, an adjustment model to address tokenizer inconsistencies, and a bug-fixing model. \mr{Toggle takes a buggy function as input and generates a complete  corrected function}. We investigate various styles of prompting to the bug fixing model to \mr{identify the most effective prompts that better utilize the inductive bias and significantly outperform others}. \textbf{Toggle} achieves the new state-of-the-art performance on the CodeXGLUE code refinement benchmark, and exhibits better and comparable performance on several other widely-used APR datasets, including Defects4J. In the Defects4J benchmark, our approach consistently ranks above other methods, achieving superior results in the Top-10, Top-30, Top-50, and Top-100 metrics. \mr{Besides examining Toggle's generalizability to unseen data, evaluating the effectiveness of various prompts, we also investigate the impact of additional contextual information such as buggy lines and code comments on bug localization, and explore the importance of the adjustment model. Our extensive experiments offer valuable insights and answers to critical research questions}.
\end{abstract}

\ccsdesc[500]{Software and its engineering~Software testing and debugging}

\keywords{Automated Bug Localization and Fix, Large Language Models}

\maketitle

\section{Introduction}
Automated program repair (APR) has become a crucial domain that aims to help developers fix bugs in software programs~\cite{apr-background-1, apr-background-2, apr-background-3}. With deep learning (DL) and large language models (LLMs) achieving great improvement on software engineering tasks, lots of DL-based and LLM-based APR techniques have been developed~\cite{jiang2023impact, xia2023automated, xia@alpharepair, cure, knod, zhu2021syntax, coconut, dlfix, tare, Chen2019sequencer, codit}. Existing techniques explore APR under two settings: with and without knowing the bug location. \mr{Under the first setting, where APR techniques are aware of the buggy lines, existing work typically trains deep learning models to replace these lines with generated fixed lines~\cite{coconut, cure, knod, zhu2021syntax, tare, xia@alpharepair, Chen2019sequencer, codit}. While this approach provides a fairer comparison of APR techniques by distinguishing the impact of bug localization from fixing, it lacks practicality in real-world scenarios}. Under the second setting, existing techniques either do bug localization simultaneously within one model or rely on spectrum-based fault localization tools (e.g., Ochiai~\cite{ochiai}) to provide bug location in line granularity. 

\mr{Recent LLM-based APR techniques have demonstrated significant improvement over DL-based techniques, benefiting from their extensive coding knowledge gained through pre-training. Yet, existing LLM-based techniques also rely on line granularity ~\cite{jiang2023impact, xia2023automated}. These methods provide the LLMs with buggy line information during prompting or fine-tuning, enabling them to generate the corresponding fixed lines. While there have been promising improvements, a comprehensive exploration into the varied applications of LLMs for bug fixing is still lacking. Specifically, there are still unanswered questions regarding the potential of LLMs for bug localization without reliance on assumptions or additional tools, the effectiveness of token-level bug location compared to line-granulated localization, and the impact of different types of prompts on the effectiveness of bug fixing.} In the subsequent sections, we dive deeper into prompting LLMs for bug fixing, utilizing LLMs for bug localization, and effectively integrating bug localization with bug fixing models.


\subsection{Prompting LLMs for Bug Fixing}
\begin{figure}[h]
    \small\centering
\includegraphics[width=.95\linewidth]{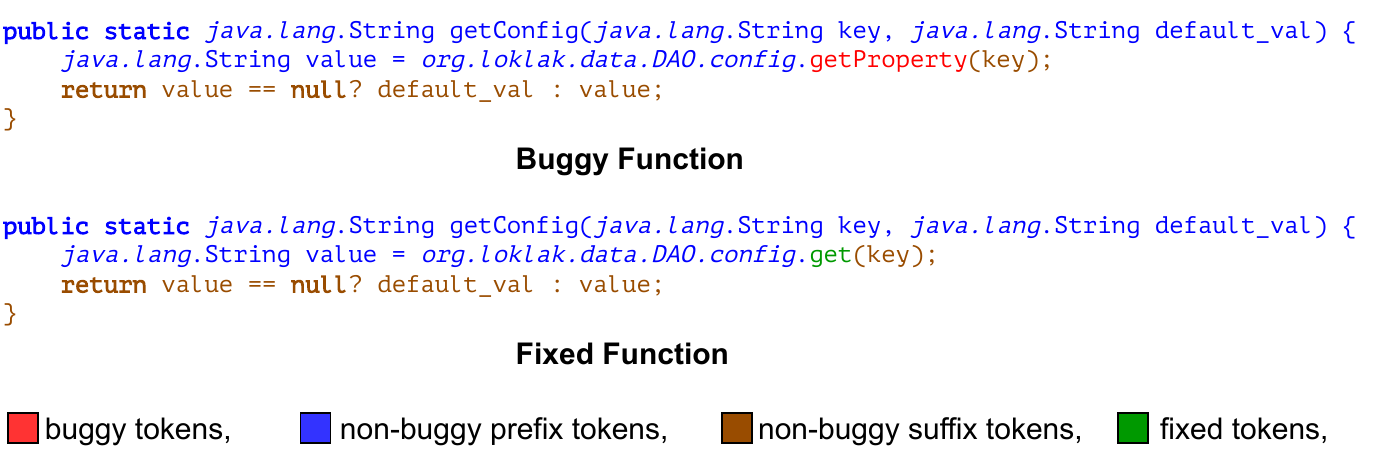}
	\caption{Example of buggy and fix code that share large non-buggy prefix and suffix}
	\label{fig:bug-fix-motivation}
	
\end{figure}

LLMs, although powerful for general code generation, are not specially designed and pre-trained for bug fixing. Thus, prompt design and fine-tuning is necessary to adjust LLMs for bug fixing task. Existing work explores the approach to input the entire buggy function to LLMs and uses comments to guide LLMs to generate the fixed function~\cite{xia2023automated}. However, it is redundant to let LLMs generate the whole function since not all the code in the original buggy function is wrong. 
\mr{It is likely that the buggy code and its corresponding fixed code share some non-buggy prefix and suffix. For example, in Figure~\ref{fig:bug-fix-motivation}, the code in blue and brown are prefix and suffix code shared by the buggy and fixed functions. Typically, only a small portion of the buggy code needs alteration. In the given example, only the ``\code{getProperty}'' needs to be replaced by ``\code{get}'' to generate the fix. Asking LLMs to generate additional code beyond this may lead to more errors and reduce the overall bug fixing capability.}

\mr{Existing method, such as~\cite{jiang2023impact}, annotates the buggy function with comments to indicate the buggy lines. LLMs can interpret these comments to identify the bug within those lines and then generate only the corresponding fixed lines}.
\mr{While this method minimizes the generation of shared prefix and suffix, it still requires generating parts of the prefix and suffix that are common to both the buggy and fixed lines. For instance, in Figure~\ref{fig:bug-fix-motivation}, ``\code{java.lang.String value = org.loklak.data.DAO.config}'' is the shared prefix, and ``\code{(key);}'' is the shared suffix. These codes are again generated by LLMs, which is unnecessary and may introduce more bugs when the required fix could be as simple as changing an identifier or operator in the line}.

\mr{To mitigate this limitation, we propose Toggle, a novel approach to token-granulated bug localization and repair, distinct from the existing line-granulated methods. Our method ensures that no shared prefix or suffix between the buggy and fixed functions will be generated. With token-granulated bug location, we explore four unique bug fixing prompts, each featuring a different arrangement of the buggy code, as illustrated in Figure \ref{fig:prompt}. These prompts are designed to prevent the regeneration of non-buggy code, eliminate redundancies in the input, compare the relative effectiveness of each prompt, and investigate further the effectiveness of different prompts in various scenarios. We fine-tune six different generative LLMs for the bug fixing task. Our findings demonstrate that preventing LLMs from generating shared prefix and suffix injects strong inductive bias, which significantly improves the bug fixing accuracy}.

\subsection{Bug Localization}
\mr{Effective bug localization is the key to bringing APR methods into practice. Bug fixing models are typically built with the awareness of the bug location, which must be accurately predicted for practical application in real-world scenarios. Most existing APR techniques primarily predict bug location at the line level, meaning they only predict that the line ``\code{java.lang.String value = org.loklak.data.DAO.config.getProperty(key);}'' contains the bug (Figure \ref{fig:bug-fix-motivation}). This line-level localization lacks finer-grained precision. Consequently, the bug fixing model still needs to figure out which specific part of the line is genuinely buggy and needs to be fixed}.

\mr{To mitigate these limitations, we propose localizing the bug at token granularity. Token-granular bug localization minimizes the generation of non-buggy shared prefix and suffix between the buggy and fixed code. Furthermore, this approach injects strong inductive bias, enabling the bug fixing model to concentrate solely on the portion requiring modification. We leverage the power of an encoder LLM (e.g., CodeT5~\cite{codet5}) with a self-attention mechanism to identify the tokens marking the beginning and end of the actual buggy code segment. Additionally, our approach leverages available contextual information (e.g., code comments) to pinpoint the buggy tokens}. In Section \ref{bug_localization}, we describe the architecture of the bug localization model and the step-by-step procedure for predicting the buggy token locations.

\subsection{Connection Between Bug Localization and Bug Fixing}
\label{sec:introduction-adjustment}
\begin{figure}[h]
    \small\centering
\includegraphics[width=.95\linewidth]{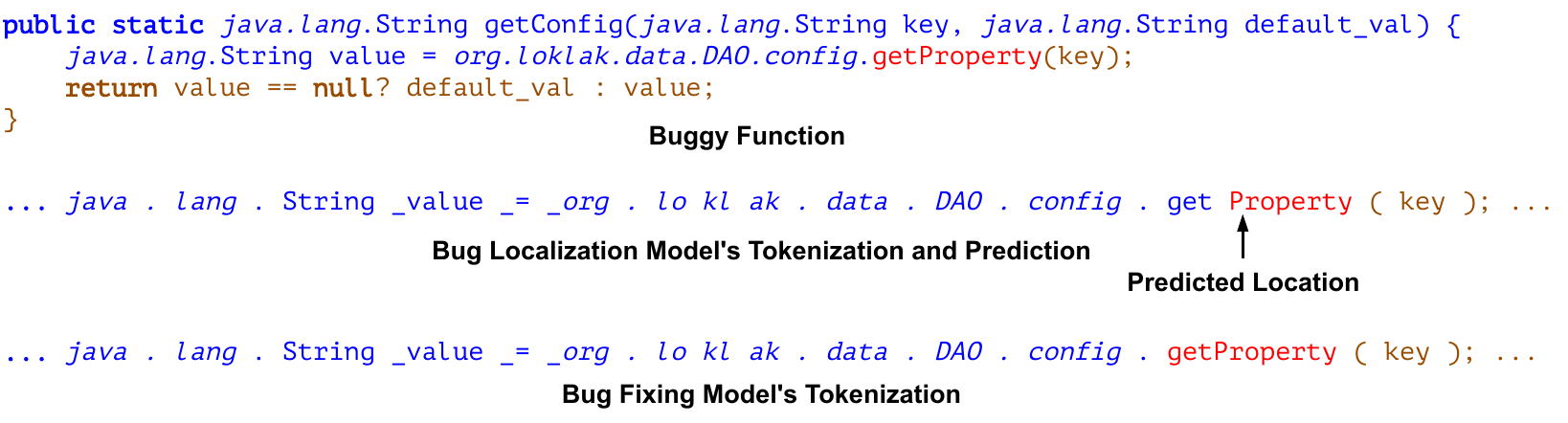}
	\caption{Discrepancy in code tokenization between the bug localization and fixing models.}
	\label{fig:adjust-model-motivation}

\end{figure}

\mr{For bug localization, we utilized an encoder-style large language model, such as CodeT5, while for bug fixing, we employed decoder-style generative LLMs, such as CodeGPT and CodeGen. One challenge with this setup is the discrepancy and inconsistency between the tokenizers of the localization and fixing models, which can impact the overall bug fixing accuracy. This issue is illustrated in Figure~\ref{fig:adjust-model-motivation}, where the correct fix involves either replacing ``\code{getProperty}'' with ``\code{get}'' or replacing ``\code{Property}'' with an empty string. The bug localization model accurately identified ``\code{Property}'' as the starting token of the buggy code}.
However, in the bug fixing model's view, ``\code{getProperty}'' is considered as one single token, thus, it is easier for the bug fixing model to replace the token ``\code{getProperty}'' with token ``\code{get}'', instead of deleting ``\code{Property}'' at the predicted bug location.

\mr{Such instances are not rare between different bug localization and fixing models, where the location predicted by the localization model is not the ideal position for the bug fixing model to generate a correct fix. To mitigate such inconsistency between tokenizers, we incorporate an optional adjustment module to bridge the bug localization and fixing models. The adjustment module is designed to find the most optimal localization considering both the localization model's prediction and the bug fixing model's performance}.

\subsection{Contributions}
To sum up, this paper makes the following contributions:
\begin{itemize}
    \item A new direction that localizes and fixes bugs at token granularity, as opposed to line granularity, thereby preventing bug fixing models from generating redundant shared prefixes or suffixes.
    
    \item A novel design of four unique prompts to fine-tune LLMs for bug fixing, demonstrating that with proper design of prompts, token-granulated localization can inject strong inductive bias and thus significantly boost the accuracy of bug fixing.
    
    \item \mr{The identification of a new challenge – inconsistencies between tokenizers in bug localization and fixing models during LLM fine-tuning – and the introduction of an adjustment module that mitigates this issue and enhances the bug fix accuracy}.
    
    \item A deep dive into the bug fixing capability of LLMs, including extensive experiments with four different prompts on six LLMs and three benchmarks. With the best framework, we developed \textbf{\toolname, \underline{To}ken-\underline{G}ranulated Bug \underline{L}ocalization and R\underline{e}pair}, which outperforms existing state-of-the-art techniques on several benchmarks.
    
\end{itemize}

\mr{In addition to introducing a new bug localization and fixing framework, we conduct a comprehensive study addressing six research questions across various topics. We investigate Toggle's effectiveness and generalizability, the influence of contextual information on bug localization, and how prompts and tokenizer inconsistencies affect bug fix accuracy. Investigating these factors across diverse datasets and conditions is crucial to confirm the effectiveness, applicability, and robustness of LLM-based methods in real-world bug fixing scenarios}. We answer the following research questions, and for each, we present the findings and their broader implications.

\begin{itemize}
    \item \mr{\textbf{RQ1} investigates the bug fixing effectiveness of Toggle, focusing on six fine-tuned LLMs from four categories across four datasets, illustrating that}
    \begin{itemize}
        \item \mr{\textbf{Finding 1:} LLMs with larger sizes yield better bug fixing accuracy after fine-tuning.}
        \item \mr{ These findings suggest selecting larger LLMs to fine-tune for bug-fixing when applying our approach if applicable.}
    \end{itemize}

    \item \mr{\textbf{RQ2} investigates Toggle's ability to generalize to unseen data, which shows that}
    \begin{itemize}
        \item \mr{\textbf{Finding 2:} The fine-tuned LLM (CodeParrot-110M) generalizes effectively to the unseen \texttt{Defects4J} benchmark~\cite{JustJE2014}, outperforming existing methods on several metrics.  }
        \item \mr{The broader implications are twofold: (1) the design of prompts during fine-tuning can make a big impact, and (2) we advocate for token-level bug location prediction as opposed to the line-level or method-level prediction prevalent in existing works.}
    \end{itemize}
    \item \mr{\textbf{RQ3} investigates the effectiveness of four different prompts on the bug fixing capability of fine-tuned LLMs, illustrating that}
    \begin{itemize}
        \item \mr{\textbf{Finding 3:} Well-designed prompts for fine-tuning result in significantly higher bug fixing accuracy. Furthermore, avoiding the generation of redundant shared prefix and suffix leads to the highest accuracy.}
        \item \mr{The broader implication is that careful prompt design and avoiding redundant code generation are critical for enhancing bug fixing effectiveness.}
     \end{itemize}
  
    \item \mr{\textbf{RQ4} investigates the effect of contextual information on bug localization, showing that}
    \begin{itemize}
        \item \mr{\textbf{Finding 4:} Additional contextual information, such as the buggy line number or code review comments, significantly boosts the accuracy of predicting both the starting and ending buggy tokens.}
        \item \mr{This finding suggests leveraging additional contextual data to assist with bug fixing, if applicable.}
    \end{itemize}
    \item \mr{\textbf{RQ5} investigates the benefits of our adjustment module, designed to address the inconsistency between the tokenizers used in bug localization and fixing models, and our result shows that}
    \begin{itemize}
        \item \mr{\textbf{Finding 5:} The adjustment module consistently improves bug-fixing accuracy on four LLMs with 110 -- 400M parameters over different datasets.}
         \item \mr{A broader implication of the finding recommends the use of the adjustment module when working with different tokenizers. As the training of the adjustment module is relatively costly, it is preferable when working with smaller LLMs.}
    \end{itemize}
     \item \mr{\textbf{RQ6} investigates, between the top-performing prompts (3 and 4), which is more effective when used with predicted bug location?}
    
    \begin{itemize}
        \item \mr{\textbf{Finding 6:} Prompt 4 outperforms prompt 3 when the predicted prefix and suffix locations are highly accurate. Otherwise, the choice between prompt 3 and 4 depends on a specific dataset, additional context, and underlying LLM backbone.}

        \item \mr{A broader implication is, when accounting for the error in bug location prediction, both prompts 3 and 4 could be optimal and require testing.
        }
    \end{itemize}
\end{itemize}

\section{IMPLEMENTATION}

In this section, we discuss six large language models and four datasets used in our study. Additionally, we provide a comprehensive overview of our bug-fixing framework, detailing its three main components: the localization model, the bug fix model, and the adjustment model.

\subsection{Large Language Models for Code}
Our method utilizes both pre-trained encoder models (for bug localization) and pre-trained auto-regressive decoder models (for bug fixing). The criteria for selecting the pre-trained models are primarily based on their popularity (frequently cited in other papers), availability (on Hugging Face), and resource constraints (compatibility with our GPU instances).

\smallskip \noindent \textbf{CodeGPT} released as a component of the CodeXGLUE ~\cite{lu2021codexglue} package, is a GPT-style model pre-trained on programming languages. It is designed to support code completion and text-to-code generation tasks. In our study, we use a checkpoint of a 12-layer, 110M parameter model that is exclusively pre-trained on Java. \mr{As it was trained on only one programming language, our experimental findings demonstrate that it does not perform as effectively as other models trained on multiple programming languages when evaluated on a multilingual dataset.}

\smallskip \noindent \textbf{CodeParrot} ~\cite{codeparrot} represents one of the earliest community versions of the pre-trained LLM on code. The complete development workflow for this model is comprehensively documented on Hugging Face, facilitating its replication by other researchers. We utilize the smaller multi-language variant of CodeParrot, which is a 12-layer, 110M parameter model. It has been pre-trained on nine programming languages: \texttt{Java}, \texttt{JavaScript}, \texttt{PHP}, \texttt{Python}, \texttt{C\#}, \texttt{C++}, \texttt{GO}, \texttt{Ruby}, and \texttt{TypeScript}.

\smallskip \noindent \textbf{CodeGen}~\cite{nijkamp2023codegen} is an auto-regressive language models designed for program synthesis. CodeGen follows the traditional auto-regressive training approach. In our research, we employ two variants of the CodeGen models, with 350M and 2B parameters, pre-trained on six programming languages. 

\smallskip \noindent \textbf{PolyCoder}~\cite{xu2022systematic}, built upon the GPT-NeoX architecture, is a recent LLM pre-trained on a substantial database containing 249GB of code spanning 12 programming languages. In our experiments, we employ the 400M and 2.7B parameter checkpoints of this model.

\smallskip \noindent \textbf{CodeT5} ~\cite{codet5} is an encoder-decoder model based on the T5 model architecture, and is specifically designed for code generation. Its encoder and decoder both utilize the transformer architecture, showing impressive results across numerous natural language tasks. While it is primarily designed for code generation, the encoder component of CodeT5 can also be leveraged for code analysis, similar to CodeBERT\cite{feng2020codebert}. In our research, we utilize the encoder component of the CodeT5-large (347M) model to predict the starting and ending positions of the buggy tokens. We choose the CodeT5-large encoder due to the limited availability of more capable, code-pre-trained encoder models. 

\subsection{Dataset}
\label{dataset}
To demonstrate the effectiveness of our method, we evaluate it on several datasets that are well-regarded in this field.

\medskip \noindent \textbf{CodeXGLUE}~\cite{lu2021codexglue} is a widely recognized machine learning benchmark for code understanding, comprising 14 datasets for a variety of code-related tasks. Our evaluation includes two datasets for code refinement from this benchmark: Tufano Small and Medium~\cite{tufano2019empirical}. The Tufano Small dataset comprises 58,350 code samples of shorter lengths, while the Tufano Medium dataset contains 65,465 code samples of medium lengths. Both datasets feature real-world buggy and fixed Java methods collected from GitHub, identified through relevant commit messages. These large-scale datasets are extensively employed in the evaluation of deep learning-based automated program repair techniques.

\medskip \noindent \textbf{CodeReviewer}~\cite{li2022codereviewer} is one of the newest and biggest datasets for numerous code related tasks, including program repair. This comprehensive dataset comprises 183,881 bug-fixing samples across various programming languages. Unlike the Tufano dataset, which focuses on code changes and bugs, the CodeReviewer dataset includes both code review comments and the associated code—before and after the review feedback is addressed. These comments provide valuable contextual information about the reasons for the changes, potentially increasing accuracy in bug localization and improving bug fixing results.



\medskip \noindent  \textbf{Defects4J}~\cite{JustJE2014} is a widely used dataset and framework for reproducible Java bugs, consisting of a total of 835 bugs from 17 open-source Java projects. For each of these bugs, the dataset provides both the buggy and the fixed versions of the code. Defects4J framework also includes unit test cases, allowing automated program repair, test case and test oracle generation methods to evaluate their effectiveness. For these reasons, Defects4J has been extensively used in software engineering research ~\cite{coconut,cure,dlfix, rewardrepair, 10.1145/3611643.3616265}. In this work, we investigate the generalizability of our fine-tuned model by generating corrective patches for Defects4J bugs.

\medskip \noindent \textbf{GitHub} ~\cite{zhu2021syntax} dataset is curated from Java projects hosted on GitHub. It includes 1,083,185 commits, predominantly identified by messages referencing terms such as `fix', `solve', `bug', `issue', `problem', and `error'. The dataset specifically focuses on patches that either modify a single statement or introduce a new one. For rigor and uniqueness, patches associated with the Defects4J project or those resembling any in Defects4J v1.2 or v2.0 were meticulously excluded, based on an AST comparative analysis. In this paper, this dataset, along with Tufano, is used for fine-tuning our models and evaluating them on the Defects4J samples.


\subsection{Overview of \toolname}

\begin{figure}[h]
    \small\centering
\includegraphics[width=.95\linewidth]{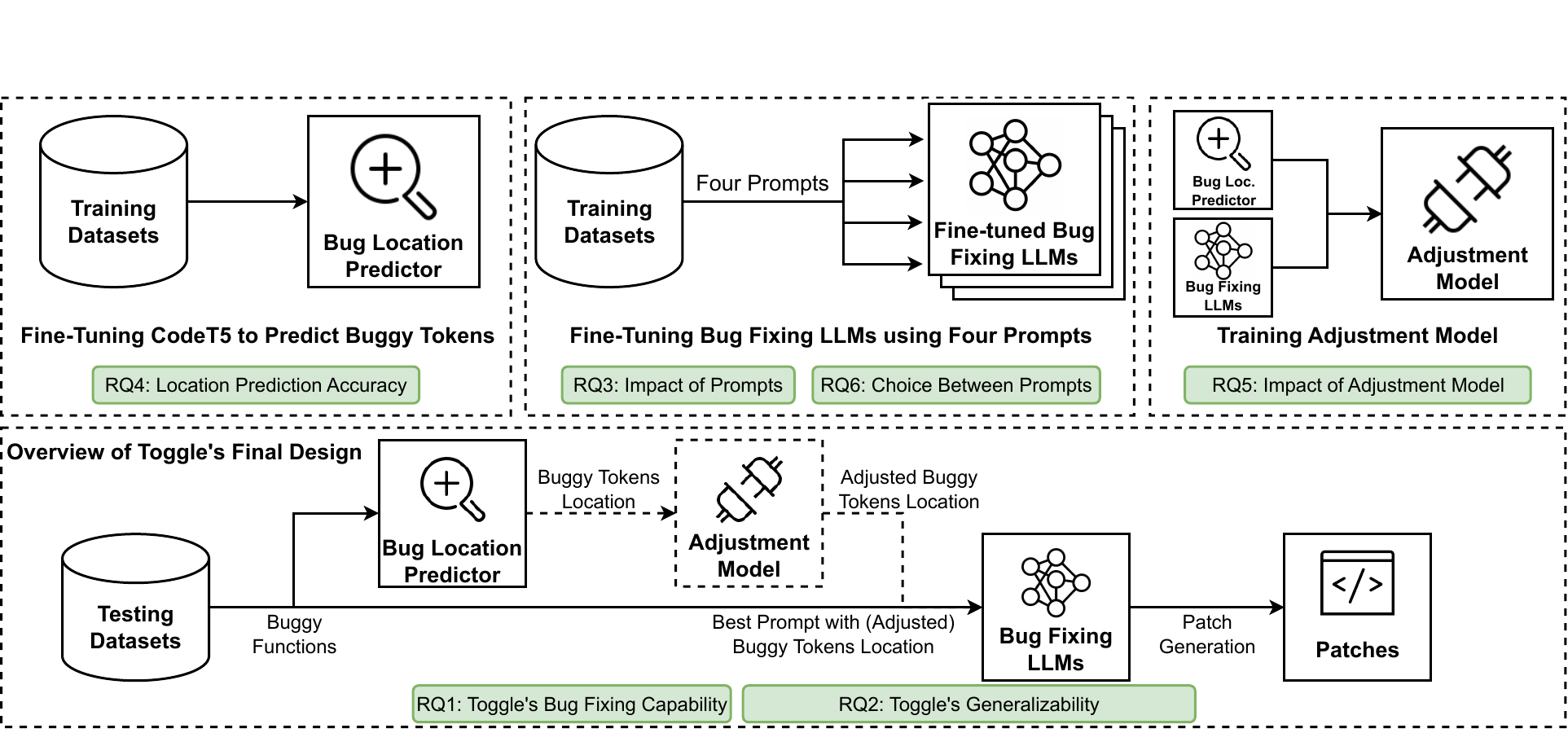}
	\caption{Overview of \toolname{} and experimental designs.}
	\label{fig:overview}
\end{figure}
\mr{Figure~\ref{fig:overview} presents the pipeline of our proposed approach, \toolname, along with a brief overview of the experimental design. Following the discussions in the previous sections, it is crucial for our framework to accommodate various prompting styles associated with different levels of utilizing inductive biases. As an end-to-end bug fixing tool, Toggle consists of three main components: bug localization model, bug fixing model, and an optional adjustment model. The bug localization model is used to predict which part of a given buggy function needs to be changed, the bug fixing model takes the buggy function as well as the predicted bug location and generates the fixed code. The adjustment model in between tunes the predicted bug location if needed, to enable the bug-fixing model to generate better patches.}

\mr{The major benefits of this partitioning design include: (1) it separates bug localization from bug fixing, simplifying the use of existing pre-trained LLMs compared to model that mixes localization and fixing; (2) it facilitates the integration of contextual information for bug localization, such as code review comments from the reviewer; and (3) it offers greater flexibility in designing various prompting styles for the bug fixing model, unlike models that handle localization and fixing simultaneously.  We discuss each component in the following sections.}

\subsubsection{Bug Localization Model: Integration of contextual information} 
\label{bug_localization}

\begin{figure}[h]
    \small\centering
\includegraphics[width=.7\linewidth]{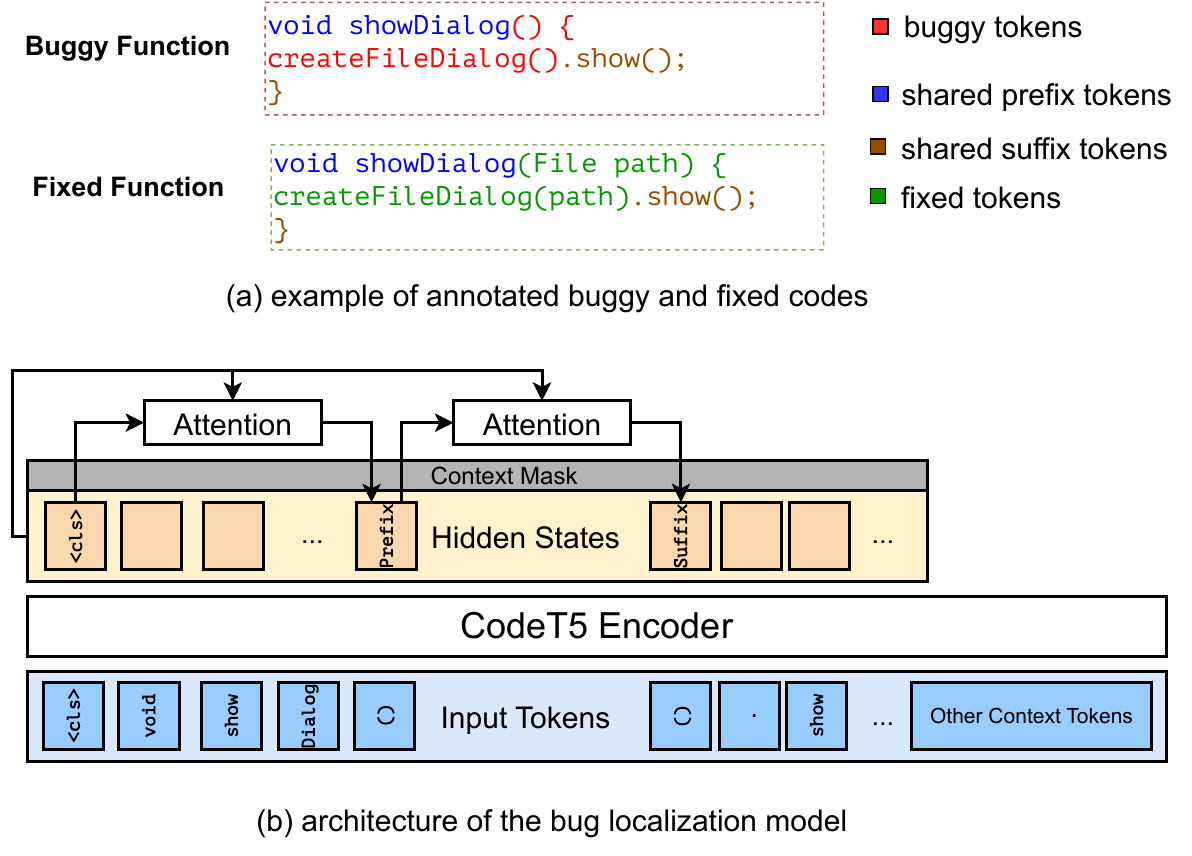}
	\caption{\mr{(a) example for buggy and fixed code , (b) architecture of the bug localization model}}
	\label{fig:localization}
\end{figure}

\mr{Figure \ref{fig:localization} (a) shows a buggy function and its corresponding fix, with colored annotations highlighting the buggy tokens, shared non-buggy prefix tokens, shared non-buggy suffix tokens, and the fixed token. We explain the bug localization steps using this example for illustration. Figure \ref{fig:localization} (b) shows the architecture of the bug localization model
which is based on the pre-trained CodeT5 encoder backbone. }

\mr{Given a buggy function, the bug localization model is trained to identify the end of shared prefix, marking the starting position of the buggy token, and the beginning of the shared suffix, indicating the end position of the buggy token. The tokens located between these two positions are considered buggy and require fixing. For example, in the buggy function shown in Figure~\ref{fig:localization} (a), the shared prefix and suffix are highlighted in blue and brown, respectively, while the buggy tokens are indicated in red. The bug localization model is trained to localize the red part, by predicting the first ``\code{()}'' as the starting buggy token and the second ``\code{()}'' before ``\code{.show}'' as the ending buggy token. With such localization, we know the tokens between these locations need to be fixed.}

In general, given the input tokenized buggy function consisting of $n$ code tokens and optional contextual information (e.g., code review comment or commit message) consisting of $m$ tokens, the input $\boldsymbol{t}=(t_{cls}, t_1, t_2,\ldots,t_n, t_{n+1}, \ldots, t_{n+m})$ is encoded by CodeT5's encoder, resulting in the embedding $\boldsymbol{e}=(e_{cls}, e_1, e_2, \ldots, e_n)$ of code tokens. \mr{For the example shown in Figure~\ref{fig:localization} (a), $e_1, e_2, \ldots, e_n$ are the embeddings of code tokens ``\code{void}'', ``\code{show}'', $\ldots$, ``\code{\}}''.}

$t_{cls}$ token is a special token added at the beginning of the input. The embedding ($e_{cls}$) of the $t_{cls}$ token serves as the query vector for performing the self-attention operation on the embeddings of the other code tokens, denoted by $\boldsymbol{e}$.


\begin{equation}
\begin{aligned}
    \boldsymbol{a}^{pre} &= \text{softmax}(W^{pre}_qe_{cls} \cdot W^{pre}_k\boldsymbol{e}) \\
    i^{pre} &= argmax_i^n\boldsymbol{a}^{pre}
\end{aligned}
\end{equation}
, where $W^{pre}_q$ and $W^{pre}_k$ are additional trainable weights and $\boldsymbol{a}^{pre}$ is the attention scores. The prefix token (i.e., starting buggy token) is predicted as the token with the highest attention score. \mr{For the example shown in Figure~\ref{fig:localization} (a), the predicted starting buggy token is ``\code{()}'' at index 4 ($i^{pre} = 4$).} For the suffix location prediction (i.e., ending buggy token), the embedding corresponding to the index of the prefix location (e.g., $e_4$ in the given example) is used as the query vector ($e_{i^{pre}}$) for self-attention. 
\begin{equation}
\begin{aligned}
    \boldsymbol{a}^{suf} &= \text{softmax}(W^{suf}_qe_{i^{pre}} \cdot W^{suf}_k\boldsymbol{e}) \\
    i^{suf} &= argmax_i^n\boldsymbol{a}^{suf}
\end{aligned}
\end{equation}
, where $\boldsymbol{a}^{suf}$ is the attention score and the suffix token is predicted as the token with the highest attention score. \mr{In the given example, the suffix token is predicted as the second ``\code{()}'' with index 8 ($i^{suf} = 8$).} Note that during training, teacher forcing technique is used, where the embedding corresponding to the ground truth prefix location is used as the query vector. In contrast, during inference, the embedding associated with the predicted prefix location is used as the query vector.

\mr{Additional contextual information can be used as the context mask, as shown in Figure~\ref{fig:localization} (b). This mask is typically employed as the masking function when calculating the loss, ensuring that locations outside the contextual window have zero probabilities of being selected. In our study, we used buggy line numbers and code reviewer comments as contextual information.
When the buggy line numbers are available, the context mask ensures only the tokens included in the buggy lines are kept in the attention process to predict the buggy location. When textual context such as reviewer comments is available, this additional textual information can be directly concatenated to the end of the code tokens to undergo the same encoding process. In this case, the context mask excludes these reviewer comments tokens in the attention process. Our experimental study in RQ4 shows that additional contextual information can significantly enhance bug localization accuracy.}

\subsubsection{Bug Fixing Model: Incorporation of inductive bias.} 
\begin{figure}[h]
    \small\centering
\includegraphics[width=\linewidth]{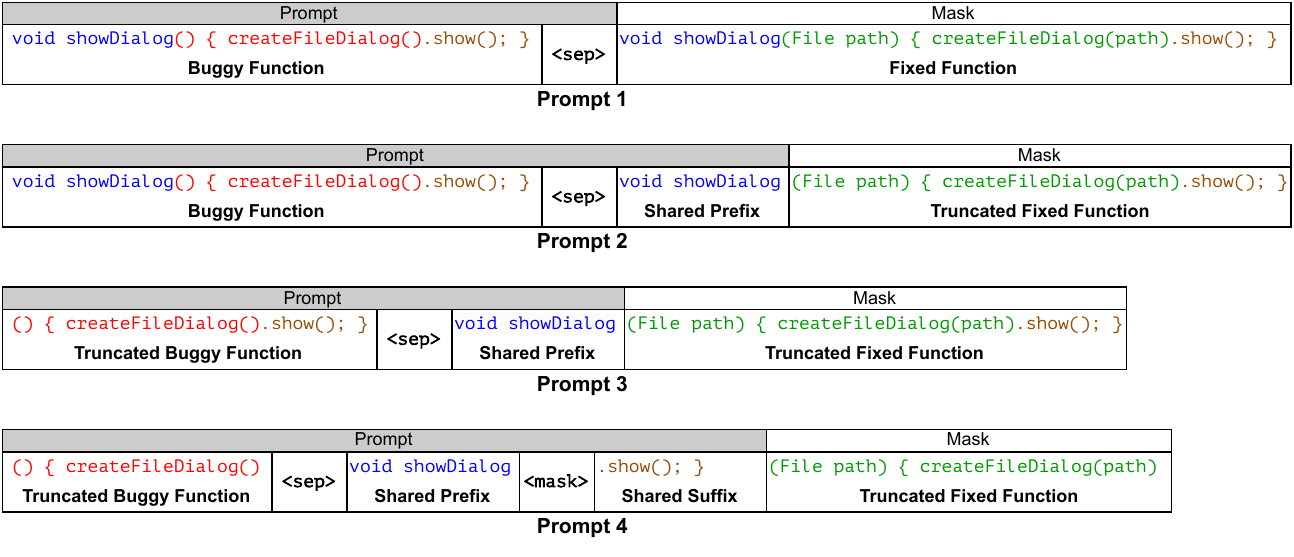}
	\caption{Four different prompts investigated. {\color{blue}Blue} tokens are non-buggy shared prefix. {\color{red}Red} tokens are buggy code. {\color{brown}Brown} tokens are non-buggy shared suffix. {\color{green}Green} tokens are fixed code.}
	\label{fig:prompt}
\end{figure}

\mr{ To enhance the performance of medium-size models and enable them to compete with larger models, it is crucial to inject `strong' indicative bias. This bias, rooted in domain-specific expertise, serves to mitigate the limitations associated with their smaller size. For bug fixing, the location of the bug is an important inductive bias. Effectively utilizing this bias can substantially enhance the bug-fixing performance of language models. This section explains how we use bug locations to create different prompts for bug fixing, shown in Figure~\ref{fig:prompt}.}

\mr{As mentioned earlier in Section 1.1 and Figure \ref{fig:bug-fix-motivation}, most bugs and their fixes typically involve only a small portion of a method. To enhance the focus of LLMs on the bug, we separate it from the non-buggy content, placing the buggy portion at the start of the prompt. This positioning ensures that LLMs concentrate on regenerating only the buggy part, minimizing the risk of introducing new bugs into the non-buggy sections. The non-buggy portion of the code is still included in the prompt so that LLMs can learn from it, allowing one to construct the entire fixed code with the generated fix.}

\textbf{Prompt 1} is a typical code completion style prompt that requires the model to generate an entire fixed function to replace the buggy function. This prompt does not utilize the bug localization bias. Therefore, one notable limitation is that it does not distinguish between the portions that need fixing and those that do not. For example, the shared prefix ``\code{\color{blue} void showDialog}'' and the shared suffix ``\code{\color{brown}.show();}'' from the buggy function part of Prompt 1 in Figure~\ref{fig:prompt} require no change.

\textbf{Prompt 2} partially leverages the location bias and 
it excludes the shared prefix tokens from prediction. The shared prefix tokens ``\code{\color{blue} void showDialog}'' are provided in the input prompt so that the bug fixing model does not need to re-generate them and only needs to generate the truncated fixed function. In prompt 2, the shared prefix appears twice: once at the start of the buggy code and again at the beginning of the fixed function. This redundancy could be optimized by truncating the buggy function, as demonstrated in prompt 3 and 4.

\textbf{Prompt 3} eliminates redundancy by truncating the shared prefix tokens from the buggy code. This approach not only prevents the model from predicting the non-buggy prefix but also indicates the model that modifications are required from the start of the prompt. By introducing such strong inductive bias, the bug fixing model is more effectively guided to start generating the fixed code from the beginning of the input. Once the model generates the truncated fixed code, appending the shared prefix enables the creation of a complete fixed code for the developer.

\textbf{Prompt 4} builds on and refines the strategy used in prompt 3. In addition to truncating the shared non-buggy prefix tokens (``\code{\color{blue} void showDialog}'') from the buggy code, it also truncates the shared non-buggy suffix tokens (``\code{\color{brown} .show();}''). This approach not only injects an inductive bias but also reduces the number of tokens the model needs to generate. Constructing prompt 4 requires two locations: the starting and ending locations of the buggy tokens. In practice, when the exact bug locations are not known, they must be predicted using our location prediction model. However, accurately predicting both locations is difficult and errors in prediction can affect the prompt's overall effectiveness. Therefore, the choice between prompt 3 and prompt 4 requires further investigation, which will be addressed in Section 3.6.

\mr{All four prompts are designed to support bugs that span multiple lines of code. For example, prompt 4 uses the format [truncated buggy function] <sep> [shared prefix] <sep> [shared suffix], where the segment within [] can encompass multiple lines of code. In cases where changes occur on non-continuous lines, we treat the entire block of code between the first and last buggy tokens as a single continuous truncated buggy section. This approach ensures that our method remains effective even when modifications are required across different lines.}

\subsubsection{Adjust Model: Connection between bug localization and bug fixing}
\mr{One disadvantage of separating the bug localization model and the bug fixing model is that they operate independently of each other. As discussed in Section~\ref{sec:introduction-adjustment}, this can cause discrepancies between the different tokenizers used by the LLMs. To mitigate such issues, we designed an optional adjustment module. This module also offers additional benefits. For example, we observe that even when the bug localization model fails to predict the exact location, as long as the predicted starting and ending buggy token locations are before and after the ground truth locations, respectively, the bug fixing model can still fix the bug. The adjust model seeks to achieve such optimum by adjusting the predicted locations so that the bug fixing model can achieve the overall best accuracy.}




\begin{figure}[h]
    \small\centering
    \includegraphics[width=0.99\linewidth]{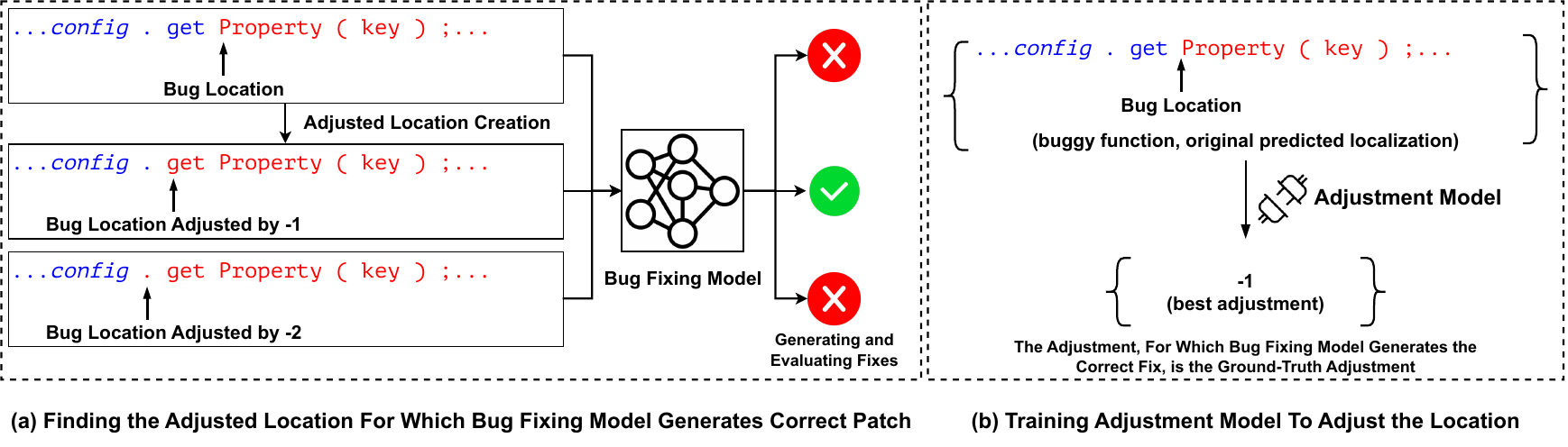}
    \caption{\mr{Overview of training the adjustment model}}
    \label{fig:adjust_overview}
\end{figure}


\mr{The adjustment model, implemented as a multinomial classification model, utilizes the CodeT5 encoder with a fully connected layer on top of the predicted location's embedding. To collect the training dataset for fine-tuning the adjustment module, we perform the following steps:}

\begin{itemize}

\item  \mr{We predict buggy token locations using the fine-tuned bug localization model. }

\item  \mr{Next, as shown in Figure \ref{fig:adjust_overview} (a), we prepare several samples for each buggy code by shifting the buggy token location around the predicted location. In our study, we shifted the buggy token location by -3 to +3. Therefore, if the predicted starting buggy token location is at \code{i}, we construct the prompts using \code{buggy\_function[i:]} as the truncated bug. With shifts, we also use \code{buggy\_function[i-1:]}, \code{buggy\_code[i+1:]} to construct truncated bug. As we use the shift range from -3 to +3, for each sample, we have a total of seven different versions of the input for the same bug.}


\item  \mr{Next, we utilize the previously fine-tuned bug fixing model to infer bug fix patches for the shifted samples prepared in step 2. We collect the shifts for which bug fixing model generates correct patches. These shifts are considered as the ground truth shift during the adjustment module training. }

\item \mr{Finally, for training the adjustment module, we prepare a dataset as \code{\{X: (buggy\_function, predicted\_loc), Y: adjustment\}}. As shown in Figure \ref{fig:adjust_overview} (b), the adjustment module takes the buggy code, predicted location as the inputs and predict the adjusted locations with which the bug fixing model should generate the correct fix. }

\item \mr{For each dataset, we utilize the validation set to learn the optimum location shift that enables the bug fixing model to generate the correct fix. For final bug fixing accuracy, we utilize the test dataset. Training larger models for location adjustment can be resource-intensive. Hence, we assess the adjustment model's performance using smaller models, such as CodeGPT-110M, CodeParrot-110M, CodeGen-350M, and PolyCoder-400M. Our experimental findings indicate that bug fixing models across the spectrum, from 110M to 400M parameters, significantly benefit from the location adjustment.}

\end{itemize}

\section{Experimental Study}


We answer the following research questions: 

\textbf{RQ1:} How does the bug fixing capability of Toggle compare to other methods?

\textbf{RQ2:} \mr{How well do the fine-tuned models generalize to unseen data?}

\textbf{RQ3:} How do different prompts affect the capability of bug fixing models to fix bugs?

\textbf{RQ4:} Does additional context along with the buggy code improve location prediction accuracy?

\textbf{RQ5:} Does the adjustment module improve bug fixing accuracy?

\textbf{RQ6:} \mr{Between the top-performing prompts (3 and 4), which one performs better with predicted bug location? }

\mr{In Figure \ref{fig:overview}, we illustrate which components were used to conduct each research study.}

\subsection{\textbf{RQ1: Toggle's bug fixing effectiveness}}

This research question investigates the automated bug-fixing effectiveness of our proposed method, Toggle. 

\subsubsection{Experimental Setup} In this study, we consider four datasets: Tufano Small, Tufano Medium, CodeReviewer w/o and w/ comments. These datasets are widely used for evaluating deep learning-based APR methods~\cite{tufano2019empirical,hu2022fix,phan2021cotext,li2022codereviewer,hong-etal-2021-fix,panthaplackel2020copy,chi2022seqtrans}. Details on the datasets are provided in Section \ref{dataset}. First, we fine-tune our location prediction model, the CodeT5-large encoder, with each of the four datasets to identify the starting location of the buggy token. Based on these predicted locations, we construct prompt 3 (\textbf{<truncated bug> [sep] <shared prefix>}) to fine-tune various bug fixing models. The datasets are divided into training, validation, and test sets, with proportions of roughly 80\%, 10\%, and 10\%, respectively. We then use the test set to generate corrective patches with our fine-tuned model and evaluate accuracy using the exact match (EM) metric. The EM metric, commonly used to evaluate deep learning-based automated program repair, considers a solution correct if it exactly matches the ground truth. Given the absence of test cases in these datasets, we consider the EM metric a more fairer measure of accuracy than either BLEU or CodeBLEU scores.

\begin{table*}[h]

\footnotesize\centering
\caption{Exact match accuracy (\%) of Toggle instances with six large language models across four datasets}\label{tab:rq1}

\begin{tabular}[t]{l|c|c|c|c}
\toprule
{\diagbox[width=11em]{\thead{\textbf{LLM Backbone}}}{\textbf{Dataset}}} & \textbf{\thead{Tufano\\ Small}} & \textbf{\thead{Tufano\\ Medium}} & \textbf{\thead{CodeReviewer\\ (w/o comment)}} & \textbf{\thead{CodeReviewer\\ (w/ comment)}} \\
\midrule
Baseline & \thead{CoText\cite{phan2021cotext}: 22.64,\\NSEdit\cite{hu2022fix}: 23.86} & \thead{NSEdit: 13.46,\\CoText: 15.36} & \thead{CodeT5\cite{li2022codereviewer}: 11.48,\\NSEdit: 11.97} & \thead{NSEdit: 23.2,\\CodeT5: 24.42} \\\hline
 
CodeGPT-110M & 21.22 & 14.42 & 8.38 & 12.26 \\
CodeParrot-110M & 21.86 & 14.22 & 10.68 & 20.04 \\
CodeGen-350M & 23.19 & \textbf{15.43} & 11.55 & 19.53 \\
PolyCoder-400M & \textbf{23.96} & \textbf{15.79} & 11.68 & 22.76 \\
CodeGen-2B & \textbf{24.73} & \textbf{16.19} & \textbf{12.31} & \textbf{25.59} \\
PolyCoder-2.7B & \textbf{25.07} & \textbf{16.19} & \textbf{12.16} & 22.92 \\
 
\bottomrule
\end{tabular}
\end{table*}

\subsubsection{Baseline} For our baselines, we selected the two best-performing models, based on our current knowledge. Given that the Tufano Small and Medium datasets are part of the CodeXGLUE benchmark~\cite{CodeXGLUE-Leaderboard}, our selection was based on the highest-ranking models from the leaderboard. For the CodeReviewer dataset, our baselines include the CodeT5 model as reported in \cite{li2022codereviewer} and NSEdit \cite{hu2022fix} model. These models achieve the top two accuracies on the CodeReviewer dataset.

The NSEdit method~\cite{hu2022fix}, a 12-layer encoder-decoder with 223M parameters, expands to around 500M including the re-ranking model. It achieved 23.86\% EM accuracy on the Tufano Small dataset and 13.46\% on the Tufano Medium. Another baseline, CoText \cite{phan2021cotext}, also has 223M parameters but supports up to 11B, achieving 22.64\% on the small and 15.36\% on the medium dataset. 

On the CodeReviewer with comment dataset, the CodeT5 model achieved an accuracy of 24.42\%, as reported in the original paper \cite{li2022codereviewer}\footnote{In the CodeReviewer paper, specialized additional pre-training increased accuracy to 30.32\%. Since the pre-training was custom and the dataset is not public, we compare our results with their reported CodeT5 results.}. 
For the CodeReviewer without comment dataset, the original paper did not report any accuracy \cite{li2022codereviewer}. However, replicating their method on this dataset is straightforward and yields an accuracy of 11.48\%. For both CodeReviewer with and without comment datasets, we run the NSEdit model~\cite{hu2022fix} and report the accuracies in Table \ref{tab:rq1}.




\subsubsection{Results}For the Tufano Small dataset, PolyCoder-400M, CodeGen-2B, and PolyCoder-2.7B exceed the baseline accuracy of 23.86\% set by NSEdit, with PolyCoder-2.7B reaching the highest accuracy of 25.07\%. On the Tufano Medium dataset, CodeGen-350M, PolyCoder-400M, CodeGen-2B, and PolyCoder-2.7B exceed CoText's baseline accuracy of 15.36\%, with PolyCoder-2.7B achieving the highest accuracy of 16.19\%.

For the CodeReviewer without comment dataset, both CodeGen-2B and PolyCoder-2.7B outperform the baseline accuracy of 11.97\% set by NSEdit. Of these, CodeGen-2B model achieves the highest accuracy of 12.31\%. For the CodeReviewer with comment dataset, CodeGen-2B achieves an accuracy of 25.59\%, outperforming the baseline accuracy of 24.41\% set by CodeT5. We observe that CodeGPT underperforms relative to all other models, primarily because CodeGPT is exclusively pre-trained on Java code, while the CodeReviewer dataset is multilingual.

\mr{Even though this study considered prompt 3, we observed that some models perform better with prompt 4, depending on the dataset and additional context. In RQ4 and RQ6, we further explore the specific conditions that lead to better outcomes for each prompt.}
\\
\finding{\mr{Toggle instances that utilize larger pre-trained LLM backbones generally lead to better performance. This aligns with our expectation as prompt 3 effectively is as a code completion task, a domain these backbones are explicitly pre-trained for. Consistently, within the same backbone family, the larger model outperforms its smaller counterpart.}}

\subsection{RQ2: Toggle's generalizability}
This research question explores Toggle's generalizability to unseen data. To this end, we applied our fine-tuned location prediction and bug fixing models to generate correct patches for Defects4J bugs, which were not included in the fine-tuning process.

\subsubsection{Experimental Setup}
\noindent\textbf{Dataset.} \mr{Defects4J is a benchmark dataset consisting of real-world Java bugs from 17 projects. This dataset is widely used for evaluating automated program repair methods. From Defects4J, we primarily focus on `single-hunk' bugs, which can be fixed with changes in a single contiguous block of code. This type of bug is well-suited for APR methods and has been extensively used in prior research~\cite{coconut,cure,dlfix,rewardrepair,knod}. This study considers a total of 240 single-hunk bugs from the Defects4J benchmark.}

\medskip\noindent\textbf{Patch generation with fine-tuned model.} \mr{For location prediction, we employ our fine-tuned CodeT5 bug localization model. As we did not fine-tune the model on the Defects4J dataset, to ensure diversity of sample distribution seen by the model, we fine-tune it on different datasets: Tufano Small, Tufano Medium, and GitHub~\cite{zhu2021syntax}. For bug fixing task, we select CodeParrot, a 110M parameter model, due to its superior performance over the similarly sized CodeGPT-110M and comparable performance to larger models like CodeGen and PolyCoder. Considering the resource-intensive nature of larger models, we used the CodeParrot model fine-tuned on same three datasets.}

\mr{To generate patches for Defects4J bugs, we first predict bug locations with three fine-tuned CodeT5 models. Then, similar to RQ1, we construct prompt 3 (\textbf{<truncated bug> [sep] <shared prefix>}), as shown in Figure \ref{fig:prompt}. Each fine-tuned bug fixing model then generates 70 patches per buggy input sample, resulting in a total of 210 patches for each bug. This quantity is relatively small compared to the number of patches generated by most existing APR techniques~\cite{coconut,cure,rewardrepair,xia@alpharepair,knod,tenure} evaluated on Defects4J.}



\medskip\noindent\textbf{Accuracy computation via test execution.} \mr{Unlike RQ1, where accuracy was determined via token-by-token exact match metric, this experiment utilizes the test cases from the Defects4J framework to validate the generated fixes. A patch is considered correct if it successfully passes all test cases. To further enhance our confidence, we manually executed all corrective patches to ensure they indeed pass all tests and effectively fix the bugs.}

\subsubsection{Results} Table \ref{tab:defects4j} presents a summary of the datasets used for fine-tuning each model and their performance in bug localization and fixing. The bug localization models accurately predicted the starting location of the buggy token for 139 bugs. In 209 cases, the models predicted the location before the actual ground truth, allowing the bug fixing models to generate corrective patches even with imperfectly predicted locations.

\begin{table*}[h]
\footnotesize\centering
\caption{Dataset used to fine-tune localization and bug fixing models, total correct location predictions and correct patches.}\label{tab:defects4j}
\begin{tabular}[t]{c|c|c|c}
  \toprule
\textbf{\thead{Location Dataset}} & \textbf{\thead{Bug Fixing Dataset}} & \textbf{\thead{Correct Location (\#)}} & \textbf{Correct Patch (\#)} \\
\midrule
   
Tufano Small & Tufano Small + Medium & 86 (19) & 59 (22) \\
Tufano Medium & Tufano Small + Medium & 85 (10) & 45 (8) \\
GitHub & GitHub & 98 (27) &  47 (16) \\
\hline
 & & Total: 139 &  Total: 82 \\
\bottomrule
\hline
\end{tabular}
\end{table*}

In the first row of Table \ref{tab:defects4j}, the fine-tuned localization model accurately predicted 86 bug locations, 19 of which were unique. The bug fixing model then generated 59 total corrective patches, 22 being unique. The second combination, despite similar overall accuracy, featured fewer unique fixes, indicating overlaps with other combinations. The third combination identified the highest number of correct locations and ranked second in both total and unique corrective patches. Utilizing three distinct combinations allowed us to leverage the diverse strengths of each dataset. Nonetheless, we consistently generated a total of 210 patches for each bug to ensure a fair comparison with previous methods.

\begin{table*}[h]
    \footnotesize \centering
    \caption{Defects4J bug fixing performance: Toggle vs. other existing methods }\label{tab:ranking}
    \begin{tabular}{l|c|c|c|c|c|c}
    \hline
        \textbf{Techniques} & \textbf{Top-10} & \textbf{Top-30} & \textbf{Top-50} & \textbf{Top-100} & \textbf{Top-200} & \textbf{$\ge$ Top-500} \\
    \hline \hline
        \textbf{CURE}~\cite{cure} & 18 & 32 & 37 & 49 & 52 & 70\\
        \textbf{RewardRepair}~\cite{rewardrepair} & 28 & 39 & 49 & 58 & 73 & 85 \\
        \textbf{Recoder}~\cite{zhu2021syntax} & 36 & 51 & 57 & 64 & \textbf{85}& - \\
        \textbf{KNOD}~\cite{knod} & 33 & 49 & 62 & 70 & 84 & 100 \\
        \textbf{Tare}~\cite{tare} & - & - & - & - & - & 109\\
        \textbf{AlphaRepair}~\cite{xia@alpharepair} & - & - & - & - & - & 110 \\
        \textbf{TENURE}~\cite{tenure} & - & - & - & - & - & \textbf{129} \\
    \hline
        \textbf{$\text{Toggle}_{\text{CodeParrot-110M}}$} & \textbf{41} & \textbf{58} & \textbf{64} & \textbf{74} & 79 & - \\
    \hline
    \end{tabular}
\end{table*}

Table~\ref{tab:ranking} shows the number of correct patches generated for Defects4J and their rankings ("-'' indicates data not reported). When generating 10, 30, 50, and 100 patches per bug, the Toggle instance with the CodeParrot-110M backbone with prompt 3 fixes more bugs than any other existing state-of-the-art techniques we are aware of. Given that producing a large number of candidate patches and executing test cases for validation is expensive and time-consuming, our approach offers better results in terms of efficiency and practicality.
\\
\finding{\mr{Toggle demonstrates strong generalization to unseen data and outperforms existing APR methods in generating corrective patches for Defects4J bugs across the Top-10, 30, 50, 70, and 100 metrics, demonstrating its generalizability, efficiency and practicality.}}

\subsection{\textbf{RQ3: Impact of prompts on bug fixing effectiveness}}
This research question investigates how different prompts affect the bug-fixing accuracy of large language models (LLMs).

\subsubsection{Experimental Setup} To solely assess the impact of our four designed prompts (shown in Figure \ref{fig:prompt}) on the bug fixing accuracy, we utilize the ground truth locations of buggy tokens instead of predicted ones. This method ensures a controlled setup, specifically designed to prevent location prediction errors from influencing the accuracy of bug fixing. We fine-tune six bug-fixing LLMs with each of the four prompts on the Tufano Small dataset \cite{tufano2019empirical} from the CodeXGLUE benchmark \cite{lu2021codexglue}. Finally, we report their test accuracy in Table \ref{tab:RQ2}.


\begin{table*}[h]
\footnotesize\centering
\caption{Impact of different prompts on bug fixing accuracy}\label{tab:RQ2}
\begin{tabular}[t]{c|c|c|c|c}
  \toprule
{\diagbox[width=12em]{\thead{\textbf{LLM Backbone}}}{\textbf{Accuracy}}} & {\thead{\textbf{Prompt 1}\\\textbf{Accuracy(\%)}}} & {\thead{\textbf{Prompt 2}\\\textbf{Accuracy(\%)}}}& {\thead{\textbf{Prompt 3}\\\textbf{Accuracy(\%)}}} & {\thead{\textbf{Prompt 4}\\\textbf{Accuracy(\%)}}}\\
\midrule
   
\textbf{CodeGPT-110M} & 16.07 & 34.19 & 44.19 & 56.98 \\
\hline
\textbf{CodeParrot-110M} & 24.5 & 53.69 & 52.18 & 63.83 \\
\hline
\textbf{CodeGen-350M} & 22 & 48.58 & 47.52 & 60.22 \\
\hline
\textbf{PolyCoder-400M} & 23.66 & 51.32 & 51.96 & 62.29 \\
\hline
\textbf{CodeGen-2B}& 23.22 & 48.99 & 48.98  & 60.13 \\
\hline
\textbf{PolyCoder-2.7B} & 23.44 & 51.25 & 50.62  & 61.62 \\
\bottomrule
\hline
\end{tabular}
\end{table*}

\subsubsection{Results} In Table ~\ref{tab:RQ2}, each row shows the bug-fixing accuracy of a specific LLM, and each column shows the accuracy achieved with each prompt formats. CodeGPT starts with a lowest accuracy of 16.07\% using prompt 1 but shows a consistent and significant improvement across prompts, reaching 56.98\% with prompt 4. CodeParrot begins at 24.5\% accuracy with prompt 1, marking the highest accuracy among all models for this prompt. Accuracy for prompts 2 and 3 remains similar; however, with prompt 4, CodeParrot reaches 63.83\%, the highest across all models for this prompt. CodeGen-350M exhibits a similar pattern, beginning at 22\% and consistently improving to achieve up to 60.22\% accuracy. PolyCoder also shows this trend, starting at 23.66\% and consistently improving to achieve an accuracy of 62.29\%. 

Bug localization at token granularity is crucial and can impact bug fixing effectiveness. We observe that with line-granular bug locations, accuracy drops below that of prompt 2 accuracy. For instance, the CodeGPT-110M model on the Tufano small dataset records an exact match accuracy of only 28.5\%.
\\
\finding{\mr{For the LLMs we have studied, there is a noticeable improvement in bug fixing accuracy from prompt 1 to prompt 4. This suggests that the latter prompts might be more effective or relevant in assisting the models with bug fixing.}}

\subsection{RQ4: Impact of contextual information on location prediction accuracy}
\label{RQ4}

In this study, we explore how contextual information affects the accuracy of predicting buggy token locations. 

\subsubsection{Experimental Setup} To address this research question, we fine-tune the CodeT5 encoder, our location prediction model, with 
prompt consisting of buggy code tokens and optional additional context. For each dataset—Tufano Small, Tufano Medium, and CodeReviewer—we perform two types of predictions for the buggy token: one for the starting location only and another for both the starting and ending locations. This process is done twice for each dataset: initially with only the buggy code included in the prompt, and subsequently with both the buggy code and additional context. The additional context for the Tufano datasets is the buggy line number, and for the CodeReviewer dataset, it is the code review comments. The accuracies are shown in Table \ref{tab:rq4}.

\begin{table*}[h]
\caption{Accuracy of buggy token location prediction with vs. without additional context}\label{tab:rq4}
\footnotesize\centering
\begin{tabular}{c|c|c}
\hline
\textbf{Dataset} & \textbf{\thead{Start token accuracy(\%)}} & \textbf{\thead{Start + end token accuracy(\%)}}\\
\hline
Tufano Small  & 39.07 & 30.81\\\hline
Tufano Small (+ Buggy Line No.)  & 60.37 & 56.07\\\hline
Tufano Medium & 26.73 & 20.16 \\\hline
Tufano Medium (+ Buggy Line No.) & 56.66 & 48.47 \\\hline
CodeReviewer  & 39.31 & 26.22 \\\hline
CodeReviewer (+ Comments) & 58.13 & 42.38\\\hline
\end{tabular}
\end{table*}

\subsubsection{Results}In Table \ref{tab:rq4}, Column 1 presents datasets with and without additional contextual information. Column 2 presents the accuracy of starting buggy token prediction, while Column 3 shows the accuracy for predicting both the starting and ending buggy tokens. The starting buggy token is needed for prompts 2 and 3, while both starting and ending tokens are needed for prompt 4, as shown in Figure \ref{fig:prompt}.

For the Tufano Small and Medium datasets, adding buggy line numbers significantly improved prediction accuracy. For Tufano Small, this addition increased starting buggy token prediction accuracy by 21\% and both token accuracy by 26\%. For Tufano Medium, it raised starting buggy token accuracy by 30\% and both token accuracy by 28\%. In the CodeReviewer dataset, including code comments increased starting buggy token prediction accuracy by 20\% and both accuracies by 16\%. These results highlight the importance of incorporating additional context to enhance buggy token location predictions.

Across all datasets, predicting both token locations yields consistently lower accuracy than predicting just the starting buggy token. RQ3 shows that bug fixing models with prompt 4, which requires both token locations, consistently outperform other prompts when the provided locations are highly accurate. This raises a question: if both token prediction accuracy is lower than that for the starting token (needed for prompts 2 and 3), which prompt is preferable? Prompt 3 or 4? This question is explored in RQ6.
\\
\finding{Incorporating additional contextual information, such as buggy line numbers or code review comments, significantly improves the accuracy of buggy token prediction across all datasets. Predicting both the starting and ending buggy token is more challenging than predicting only the starting token.}

\subsection{RQ5: Impact of adjustment module on bug fixing accuracy}
\mr{Section 2.3.3 discusses how tokenizer discrepancies between the bug localization and fixing model can impact accuracy. To address this, we introduce an optional adjustment module, discussed in Section 2.3.3. This research question explores the adjustment module's potential to improve bug fixing accuracy. }

\subsubsection{Experimental Setup} \mr{For this experiment, we consider four datasets: Tufano Small, Tufano Medium, CodeReviewer without comments, and CodeReviewer with comments. We evaluate four models: CodeGPT-110M, CodeParrot-110M, CodeGen-350M, and PolyCoder-400M. In total, we have conducted 16 sets of experiments. For each dataset and LLM combination, we compute the bug-fixing accuracy with the adjustment module enabled and disabled. We present the results in Table \ref{tab:rq5}.}

\begin{table*}[h]
\small\centering
\caption{Bug fixing accuracy improvement due to location adjustment}\label{tab:rq5}
\resizebox{\columnwidth}{!}{%
\begin{tabular}[t]{l|c|c|c|c|c|c|c|c}
\toprule
\multirow{2}{*}  {\diagbox[width=8em]{\thead{\textbf{Dataset}}}{\thead{\textbf{LLM Backbone}}}} & \multicolumn{2}{c|}{\textbf{CodeGPT-110M}} & \multicolumn{2}{c|}{\textbf{CodeParrot-110M}} & \multicolumn{2}{c|}{\textbf{CodeGen-350M}} & \multicolumn{2}{c}{\textbf{PolyCoder-400M}}\\
\cline{2-3} \cline{4-5} \cline{6-7} \cline{8-9} 
& \textbf{\thead{Accuracy(\%)}} & \textbf{\thead{Adjusted \\Accuracy(\%)}}& \textbf{\thead{Accuracy(\%)}} & \textbf{\thead{Adjusted \\Accuracy(\%)}} &
\textbf{\thead{Accuracy(\%)}} & \textbf{\thead{Adjusted \\Accuracy(\%)}}&
\textbf{\thead{Accuracy(\%)}} & \textbf{\thead{Adjusted \\Accuracy(\%)}}\\
\midrule
Tufano Small & 21.22 & 21.97 & 21.78 & 23.51 & 23.19 & 23.63 & 23.96 & 24.75\\
\hline
Tufano Medium & 14.42 & 14.74 & 14.22 & 15.14 & 15.43 & 15.60 & 15.79 & 16.04\\
\hline
\mr{\thead{CodeReviewer\\ (-Comment)}} & \mr{8.38} & \mr{8.57} & \mr{10.68} & \mr{11.57} & \mr{11.55} & \mr{11.70} & \mr{11.68} & \mr{11.75} \\
\hline
\mr{\thead{CodeReviewer \\(+Comments)}} & \mr{12.26} & \mr{13.52} & \mr{20.04} & \mr{21.37} & \mr{19.53} & \mr{21.55} & \mr{22.76} & \mr{23.92} \\
\bottomrule
\end{tabular}
}
\end{table*}

\subsubsection{Results}\mr{In Table \ref{tab:rq5}, across all scenarios, we observe a consistent improvement in accuracy when the adjustment module is enabled, suggesting the effectiveness of the module in addressing tokenizer discrepancies.}

\begin{figure}[h]
    \small\centering
    \includegraphics[width=\linewidth]{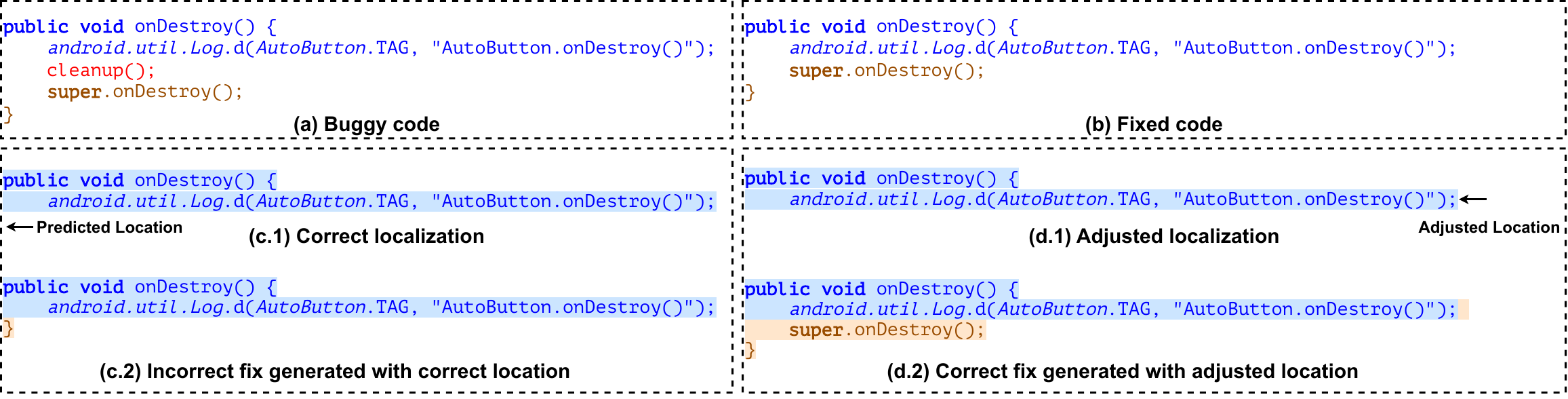}
    \caption{Example of incorrect fix (c) with correct location and correct fix (d) with adjusted location}
    \label{fig:adjust_model_example}
\end{figure}

\mr{Previously, in Figure \ref{fig:adjust-model-motivation}, we demonstrate the adjustment module's ability to address tokenizer discrepancies. In Figure \ref{fig:adjust_model_example}, we present another motivating example (new scenario) from the Tufano Small dataset (147th sample). (a) shows the buggy code, while (b) shows the fixed code which removes the erroneous code ``\code{cleanup()}''. Consequently, the correct starting buggy token location is right after the newline (``$\backslash$n'') of code ``\code{...AutoButton.onDestroy()");}''. Yet, even with the correctly predicted location, the bug fixing model prematurely closes the entire method. In Figure~\ref{fig:adjust_model_example} (c.2), the blue-background code represents the shared prefix in prompt 3, while the orange-background code indicates the model's incorrect fix.}

\mr{We find that shifting the predicted location to the left by one character—excluding the newline—causes the bug fixing model to generate a newline as the initial fix, followed by the correct fix. While the exact reason for this behavior is unclear, similar patterns frequently occur in our datasets. When a prompt concludes with an extended code line and a newline token, the bug-fixing model may prematurely end the method. This may be due to the model's exposure to similar samples during pre-training or fine-tuning. Our adjustment module identifies this pattern and accordingly adjusts the predicted location.}
\\
\finding{\mr{The buggy token location adjusted by the adjustment module consistently improves the bug-fixing capability of four LLMs with 110 -- 400M parameters, and this trend is consistent across all of the studied datasets.}}

\subsection{\textbf{RQ6: Effectiveness of prompt 3 vs 4 with predicted locations}}
\mr{Prompt 3 requires only the starting buggy token location, whereas prompt 4 requires both the starting and ending buggy token locations. In RQ3, we observe that prompt 4 outperforms all others when the provided locations are highly accurate. Furthermore, in RQ4, we notice that the prediction accuracy of both the starting and ending buggy tokens is lower than predicting only the starting token. As inaccurate locations can affect the overall bug-fixing performance, which prompt to use with predicted locations? Prompt 3 or 4?} We investigate this research question in RQ6.

\subsubsection{Experimental Setup} \mr{For each of the four datasets, we calculate four distinct accuracies for buggy token predictions: 1) starting token only, 2) ending token only, 3) partial (where the predicted starting token location precedes the ground truth and the ending token follows the ground truth), and 4) both tokens. Using the predicted locations, we employ four bug fixing models: CodeGPT, CodeParrot, CodeGen-350M, and PolyCoder-400M. For each model, we calculate the accuracy for both prompt 3 and prompt 4. }

\begin{table*}[h]
\caption{Four distinct accuracies for buggy token location prediction }\label{tab:prefix_dual_location}
\footnotesize\centering
\begin{tabular}{c|c|c|c|c}
\hline
\textbf{Dataset} & \textbf{\thead{Starting token \\only(\%)}} & \textbf{\thead{Ending token \\only(\%)}} & \textbf{\thead{Both token(\%)}} & \textbf{\thead{Partial(\%)}} \\
\hline
Tufano Small & 38.66 & 38.33 & 30.51 & 53.23\\\hline
Tufano Medium & 26.66 & 25.04 & 19.54 & 38.23\\\hline
\thead{CodeReviewer (-Comment)} & 40.56 & 37.12 & 26.22 & 60.46\\\hline
\thead{CodeReviewer (+Comment)} & 57.54 & 53.19 & 42.38 & 65.76\\
\hline
\end{tabular}
\end{table*}

\subsubsection{Results}Table \ref{tab:prefix_dual_location} presents four distinct location prediction accuracies. Accuracies for correctly predicting both the starting and ending tokens are lower than those for single token predictions, highlighting the challenge of dual token prediction. Partial accuracy suggests the bug fixing model may still generate the correct patch since the predicted tokens encompass the buggy portion.

\begin{table*}[h]
\footnotesize\centering
\caption{Bug fixing accuracy of prompt 3 and 4, with predicted buggy token location}\label{tab:prefix_dual_overall}
 \resizebox{\columnwidth}{!}{%
\begin{tabular}[t]{l|c|c|c|c|c|c|c|c}
  \toprule
\multirow{2}{*}  {\diagbox[width=8em]{\thead{\textbf{Dataset}}}{\textbf{Code Model}}} & \multicolumn{2}{c|}{\textbf{CodeGPT}} & \multicolumn{2}{c|}{\textbf{CodeParrot}} & \multicolumn{2}{c|}{\textbf{CodeGen-350M}} & \multicolumn{2}{c}{\textbf{PolyCoder-400M}} \\
\cline{2-3} \cline{4-5}\cline{6-7}\cline{8-9}
& \textbf{\thead{Prompt 3\\(\%)}} & \textbf{\thead{Prompt 4\\(\%)}}& \textbf{\thead{Prompt 3\\(\%)}} & \textbf{\thead{Prompt 4\\(\%)}} & \textbf{\thead{Prompt 3\\(\%)}} & \textbf{\thead{Prompt 4\\(\%)}} &
\textbf{\thead{Prompt 3\\(\%)}} & \textbf{\thead{Prompt 4\\(\%)}}  \\
\midrule
   
Tufano Small & 21.22 & 17.43 & 21.78 & 19.83 & 23.19 & 16.47 & 23.96 & 18.71 \\
\hline
Tufano Medium & 14.42 & 12.17 & 14.22 & 13.41 & 15.43 & 10.25 & 15.79 & 11.06 \\
\hline
\thead{CodeReviewer\\ (-Comment)} & 8.38 & 10.88 & 10.68 & 12.13 & 11.55 & 11.43 & 11.68 & 11.33 \\
\hline
\thead{CodeReviewer\\ (+Comment)} & 12.26 & 19.7 & 20.04 & 24.65 & 17.83 & 23.26 & 22.76 & 23.86 \\
\bottomrule
\hline
\end{tabular}
}
\end{table*}

\mr{Table \ref{tab:prefix_dual_overall} presents the bug-fixing accuracies for prompts 3 and 4 across various models and datasets. For the Tufano small/medium datasets, prompt 3 consistently yields better accuracy than prompt 4 in all cases. However, for the CodeReviewer datasets, prompt 4 is generally more effective than prompt 3 in most instances, which could be attributed to two main reasons. Firstly, the partial location accuracies shown in Table \ref{tab:prefix_dual_location} are much higher for the CodeReviewer datasets than for the Tufano datasets. This higher partial accuracy indicates that the predicted starting buggy token location is before the actual starting location, and the ending buggy token location is after the actual ending location, allowing the bug-fixing models a better chance to make correct predictions. Secondly, with code reviewer comments, the bug-fixing model has more context, enabling more accurate predictions. Since prompt 4 requires both locations, the higher partial accuracies and additional context from code reviewer comments mean that most models perform better with prompt 4 for the CodeReviewer datasets. In summary, the performance of prompts 3 and 4 can vary depending on the dataset and the underlying backbone LLMs, highlighting the necessity of testing both prompts to achieve optimal performance.}
\\



\finding{\mr{With predicted bug locations, both prompts 3 and 4 can perform better, depending on the datasets, the presence of additional contexts, and the underlying backbone LLMs. Generally, prompt 4 works better when the predicted starting and ending buggy token locations are highly accurate; otherwise, it is necessary to test both prompts for optimal performance.}}

\section{Related Work}
\subsection{Deep Learning-Based Automated Program Repair}

Research in deep learning-based program repair has primarily focused on three approaches: neural machine translation (NMT)-based, edit-based, and the more recent large language model (LLM)-based methods. NMT-based approaches treat the correction of code as a translation task, where the `fixed' code is translation target. This method is relatively simple to implement because both the input (buggy code) and the target (fixed code) are sequences of code, enabling direct modeling as a sequence-to-sequence (Seq2Seq) task~\cite{Chen2019sequencer,coconut,cure,rewardrepair}. 
However, a potential downside of directly predicting the fixed code is that it may encourage the models to learn the copying behavior due to the typical large overlap between the buggy and fixed code. This copying behavior causes over-fitting and overlooks the goal of editing to change the code and fix the bug.


Editing-based approaches try to directly target the portion of the code that contains bug and usually contain an explicit mechanism for bug localization. Although edits are often shorter than the fixed code, most of the existing approaches often require multiple stages of editing with complicated models~\cite{hashimoto2018retrieve, zhu2021syntax, yin2018learning, tarlow2020learning, knod, tare}. They may also rely on a graph representation of the buggy code \cite{yao2021learning, Dinella2020}, with the implicit assumption that the buggy code can be parsed into a graph, which may not be true if bugs contain syntactical errors. 

Recent studies have explored large language models (LLMs) for automated program repair (APR), showing that LLMs, either as-is or with straightforward fine-tuning, have good fixing capabilities~\cite{jiang2023impact,xia2023automated}. However, these studies did not address bug localization or its integration with bug fixing, nor did they conduct a comprehensive study on the design of prompts for fine-tuning LLMs.

\subsection{Large Language Models For Software Engineering}
LLMs have been applied to several software engineering tasks in addition to APR, including code generation~\cite{related-generation-1,related-generation-2,related-generation-3,related-generation-4}, vulnerability detection and fixing~\cite{related-vul-1,related-vul-2,related-vul-3}, fuzzing~\cite{related-fuzz} and software specification generation~\cite{related-specification}. Such adoption of LLMs typically requires fine-tuning if a decent amount of data is available, or prompt design if the prompt engineering approach is utilized.

In this work, we focus on APR only. However, the insight that better prompts lead to significantly improved fine-tuning results is widely applicable, potentially benefiting other software engineering tasks that use large language models (LLMs).

\section{Threats to Validity}

In our research, we have investigated the effectiveness of our bug fixing framework using several open-source, widely recognized datasets such as CodeXGLUE ~\cite{lu2021codexglue} and CodeReviewer ~\cite{li2022codereviewer}. These datasets present a diverse set of bugs collected from GitHub and span various programming languages. However, there's a possibility that our results may not generalize across other datasets. To address this threat and ensure broader applicability, we further evaluated the performance of our fine-tuned model on the unseen dataset, Defects4J, which is a widely used real-world bug dataset.

In our large-scale experiments, we have developed various tools and scripts to facilitate our experiments, and it is conceivable that they might contain bugs. To mitigate this threat, we have used widely used PyTorch and Hugging Face libraries. 
Additionally, we have carried out thorough validity checks and repeated each experiment several times to confirm consistency. 

In our study, we measured bug fixing accuracy using the `exact match' metric, widely adopted by the research community. Specifically for the Defects4J dataset, we determined accuracy through test validation. This involved executing the test cases on the generated patches; a patch was deemed correct if it passed all test cases. We are confident that this approach provides a reasonable and reliable accuracy metric.

\section{Conclusion}

In this study, we have explored the intricacies of LLMs in the realm of automated program repair.
To sum up, this paper makes the following contributions:
(1) Granularity Shift: We introduced a new direction that localizes and fixes bugs at token granularity instead of the traditional line granularity. This innovation drastically reduces the redundancy for bug-fixing models to produce shared prefixes or suffixes.
(2) Prompting Strategy: We presented a novel design of four diverse prompts to optimize LLMs as bug-fixing models. This approach proves that with the right prompt design, token-granulated localization brings a potent inductive bias, leading to substantial improvements in bug-fixing accuracy.
(3) Harmonizing Discrepancies: We addressed the unique challenge posed by inconsistencies between bug localization and fixing models during LLM fine-tuning. To reconcile this, we proposed an intermediate adjustment module, further strengthening the capability of our bug-fixing models.
(4) In-depth Exploration: Our extensive research into LLMs' bug-fixing capabilities encompassed experiments with four different prompts across six LLMs and five benchmarks. The culmination of this research resulted in the creation of \textbf{Toggle}, our Token-Granulated Bug Localization and Repair framework. \textbf{Toggle} has surpassed existing state-of-the-art techniques across several benchmark assessments.

Our proposed framework, \textbf{Toggle} demonstrates the success of our strategy, establishing a new benchmark in CodeXGLUE code refinement and showing notable performance on several datasets, including Defects4J. We answered several research questions, offering in-depth insights, discussions, and conclusions through meticulous experimentation. Our study not only advanced the current state-of-the-art in APR but has also paved the way for future research to explore newer dimensions of LLMs in software engineering. As the potential of LLMs continues to unfold, we anticipate Toggle's methodology to inspire subsequent models, fostering even more robust, reliable, and efficient automated software engineering solutions.


\bibliographystyle{ACM-Reference-Format}
\bibliography{bugfix-fse-24}

\end{document}